# A novel particle-in-well technology for single-molecule sequencing by surface-enhanced Raman spectroscopy

Eva Bozo [1], Pei-Lin Xin [1,3], Yingqi Zhao [1,3], Mulusew W. Yaltaye [1,2,3], Aliaksandr Hubarevich [4],

Viktorija Pankratova [5], Shubo Wang [5], Jian-An Huang [1,2,3],*

[1] Research Unit of Health Sciences and Technology, Faculty of Medicine, University of Oulu, Aapistie 5 A, 90220 Oulu, Finland.

[2] Research Unit of Disease Networks, Faculty of Biochemistry and Molecular Medicine, University of Oulu, Aapistie 5 A, 90220 Oulu, Finland.

[3] Biocenter Oulu, University of Oulu, Aapistie 5 A, 90220 Oulu, Finland.

[4] Laboratory for Information Display and Processing Units, Belarusian State University of Informatics and Radioelectronics, 6 P. Brovki, 220013 Minsk, Belarus

[5] Nano and Molecular Systems Research Unit, Faculty of Science, University of Oulu, Pentti Kaiteran katu 1, 90570 Oulu, Finland

*Email: jianan.huang@oulu.fi

## Abstract

Single-molecule surface-enhanced Raman spectroscopy based on a particle trapped in a plasmonic nanopores provides a unique method for continued and controlled detection of peptide and DNA oligonucleotides in liquid medium. However, the Brownian motion of the particle and the molecule diffusion acting on the particle hinder single-molecule sequencing. In this study, we developed a method for trapping a gold nanoparticle in an air-filled gold nanowell (particle-in-well) to stabilize the particle and provide a powerful platform for continuous single molecule readout. The unlimited resident time of the particle-in-well device with single-molecule level sensitivity elevates nucleobase detection to a new level. We present a technique capable of detecting and monitoring solid-phase molecule diffusion within the plasmonic hotspot. Furthermore, the measured spectra were employed as input data for the validation of the plasmonic hotspot size and, consequently, the distance between the particle and the well. The obtained results form the statistical and experimental base for molecular translocation and DNA sequencing technologies.

## Introduction

The techniques used for DNA and protein sequencing have been rapidly evolving due to the demands of scientific, medical, and therapeutic implications. [1] The currently available pioneer nanopore-based technologies are capable of DNA sequencing, but typically exhibit a high error rate and lack the sufficient sensitivity for amino acid discrimination. [2] Surface-enhanced Raman spectroscopy (SERS) embedded in plasmonic nanopores and nanocavities has increased the discrimination power and detection limits. In particular, the emerged intense optical field, designated as a plasmonic hot spot, boosts the Raman intensity by a factor of 100 million, making single molecule SERS detection feasible. Single-molecule SERS detection and discrimination have been reported in various powerful technical solution, including the combinations of nanoparticles and nanocavities or tip-enhanced Raman spectroscopy (TERS). [4] Our group has demonstrated the ability to control the electro-plasmonic trapping of nanoparticle by using optical tweezers, resulting in a long enough residence time for single-molecule oligonucleotide detection and the discrimination of 20 amino acids. Furthermore, the obtained spectra were applied to demonstrating peptide sequencing in two model compounds: vasopressin and oxytocin. [5] [6]

While the nanoparticle trapping technologies are applied to single molecule sequencing in liquid media, they face challenges that both the dynamics of trapped nanoparticle and molecules on the nanoparticle affect the hot spot and SERS signals. For example, the trapped nanoparticle experienced Brownian movement such that the excitation area on the molecule by the hot spot was continuously changing. Meanwhile, the molecule itself had Brownian diffusion on the nanoparticle surface. To do sequencing, the nanoparticle should be fixed so that only molecule dynamics in the hot spot would be observed by SERS. Such case is similar to TERS that enables single-molecule detection and imaging in vacuum on solid surfaces with high spatial resolution. [7] However, the stable and accurate molecular alignment remains a challenge in TERS, though the short dwell time of the molecule on the surface is not a limiting factor. For example, overlapping DNA strands hinder the recognition power, while molecular adsorption/desorption occur in the proximity of the tip, resulting in molecular dragging and detection errors. [8] Furthermore, TERS measurement requires ultra-high vacuum and low temperature to perform high-quality experiments, which result in low throughput. For example, TERS has been applied to unravel adsorbate-substrate interactions on immobilized molecules at 19K. To increase the sequencing throughput, routine measurements should be typically performed at room temperature, which requires a new tool to characterize the molecular structure and behaviour. [9]

Herein, we present a novel plasmonic particle-in-well device that capable to discriminate single-molecule by entrapping and detecting discrete single bases from oligonucleotide fragments. The nanowell array creates an intense hot spot and serves as a cage for the analyte adsorbed on a single gold nanoparticle. The gold nanoparticle, carrying the submonolayer analytes, are entrapped into the gold nanowell system via directed assembly method. Once the nanoparticle/analyte is secured at the edge of the nanowell, the resulting nanogap between the nanoparticle and nanocavity wall enhances laser excitation to generate a gap-mode plasmonic hot spot for single molecule SERS detection. Various forces, such as Brownian motion, molecular diffusion, the plasmonic force from the laser excitation, and the entropic forces, cause the oligonucleotide chains to move, vibrate, diffuse, adsorb and desorb. These motions are monitored under the hot spot, providing insights into (1) molecular diffusion speed (2) range of molecule random movement (3) Raman peak characteristics as input data for qualitative nucleobase determination.

## Results and discussion

### Particle-in-well device.

The nanowell arrays with 200 nm holes were fabricated using FIB on 100 nm thick gold film deposited on Si3N4 support wafer (1.5-2 cm x 1 cm). The molecules (peptide, oligonucleotides) were adsorbed on the surface of gold nanoparticles and directed into the nanowells by directed capillary assembly method. The nanoparticles, with the adsorbed molecules, are permanently trapped in the nanowell. The resulting plasmonic hotspot between the nanoparticle and the nanowell provides a strong electromagnetic enhancement for single molecule SERS detection (Figure 1).

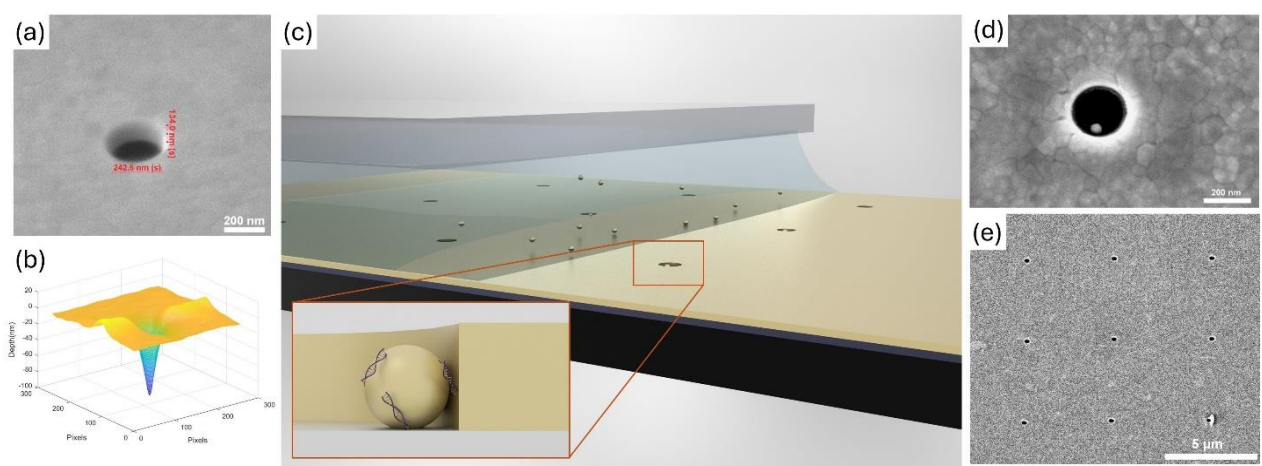

Figure 1: Directed capillary assembly for particle-in-well device for single-molecule SERS. **a** FIB drilled nanowell on 100 nm Au layer on $Si_3N_4$ wafer **b** the depth of the nanowell characterized by AFM **c** Schematic of the directed capillary assembly of gold nanoparticle, showing an entrapped gold nanoparticle with adsorbed oligonucleotide in the insert **d** SEM image of particle-in-well **e** SEM image of nanowell arrays spaced more than 4 µm apart

### Single-molecule SERS validation

To validate single-molecule SERS, a peptide fragment (Alpha-2-HS-glycoprotein 2 (FETUA), position: 132-144) was adsorbed on the gold nanoparticle surface. The peptide chain consists of the following amino acids: Cys-Asp-Ser-Ser-Pro-Asp-Ser-Ala-Glu-Asp-Val-Arg-Lys. [13] The distribution of the detected molecules depends on the size and the location of the plasmonic hotspot. If the hotspot size is comparable to that of the single amino acid, there is likelihood of detecting just one amino acid, resulting in dominant spectra showing single-molecule events. However, if the two characteristics peaks appear simultaneously, the hotspot coverage area is likely larger, resulting in multi-molecule events. The Raman peaks of the first detected spectra are 926 cm⁻

[1], 962 cm[-1], 1055 cm[-1] and 1116 cm[-1], which align with the characteristics peaks of valine ( 948 cm[-1], 965 cm[-1], 1066 cm[-1] and 1126 cm[-1]) as shown on Fig.2. The neighbouring amino acids to valine are arginine and aspartic acid. The characteristic peaks of solid arginine are at 926 cm[-1], 985 cm[-1], 1036 cm[-1], 1083 cm[-1], 1100 cm[-1] and for aspartic acid 939 cm[-1], 992 cm[-1], 1084 cm[-1], 1121 cm[-1]. The arginine and aspartic acid peaks do not overlap with the detected valine peaks, but it is worth mentioning that the peak at 926 cm[-1] could not be assigned to valine and might come from the neighbouring arginine. [14]

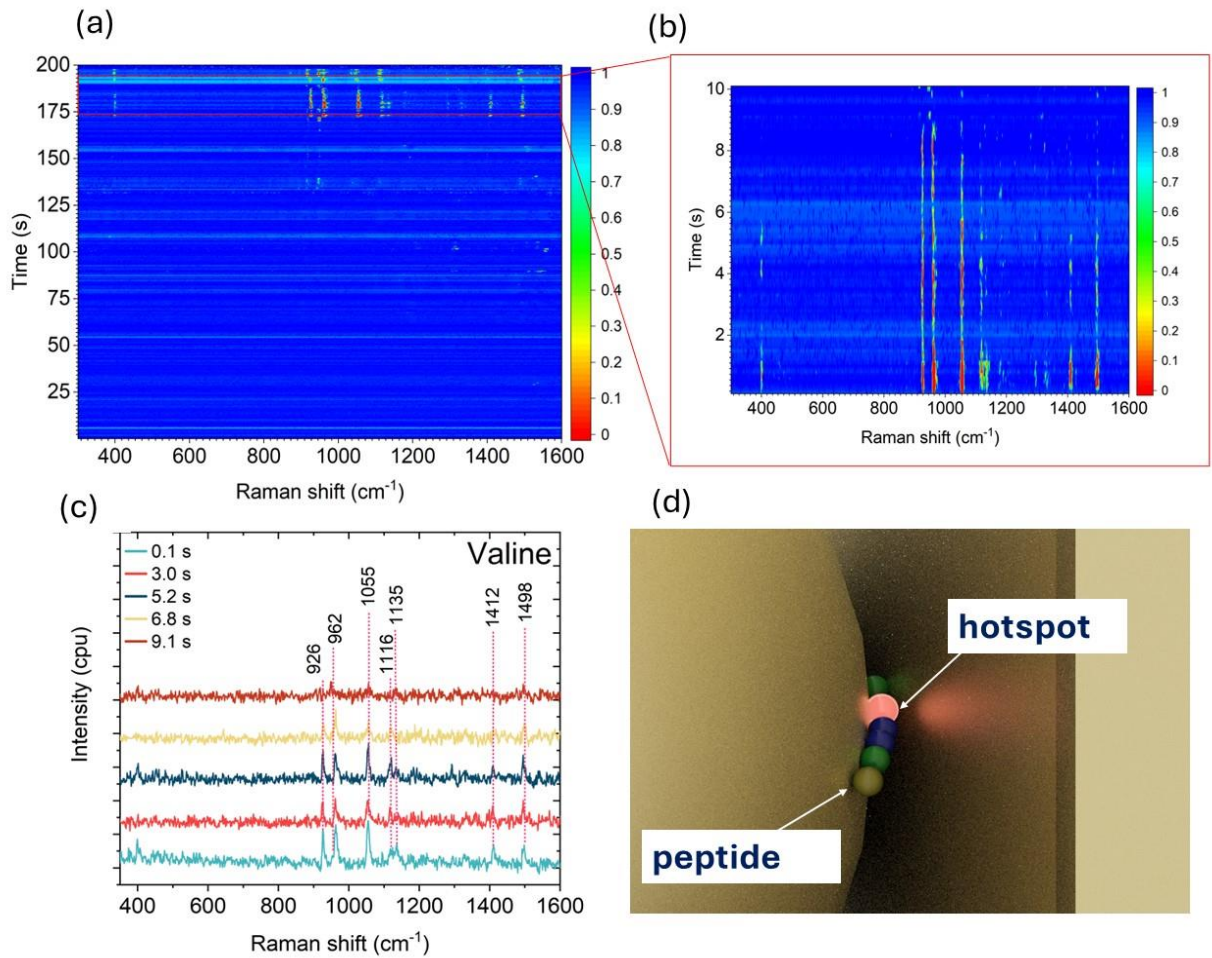

Figure 2: SERS of single molecule valine detection in peptide (FETUA) **a** Raman spectral map of FETUA (2000 spectra) **b** Extracted spectral map of the region 179.0-189.0 s **c** Peak assignment of valine **d** Schematic of peptide trapping under the plasmonic hotspot

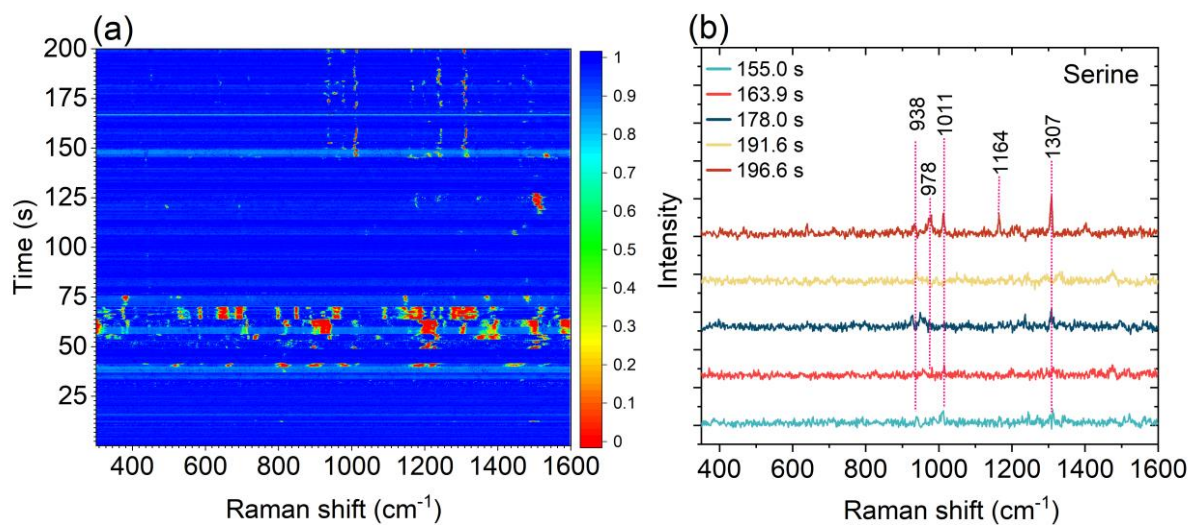

Figure 3. SERS of single molecule valine detection in peptide (FETUA) **a** Raman spectral map of FETUA (2000 spectra) **b** Peak assignment of serine

To further validate our findings, we conducted a second experiment that is aligned with the initial measurement but provides stronger evidence. In the second experiment the conditions were the same. From a dataset of 2000 spectra (Fig3.a), 50 was extracted, which exhibited the most characteristic peaks with minimal fluctuations. Following this, we performed peak assignment using 5 of these spectra and identified 5 Raman shifts withing this range (938 cm$^{-1}$, 978 cm$^{-1}$, 1011 cm$^{-1}$, 1164 cm$^{-1}$, 1307 cm$^{-1}$). Comparing these shift to the literature, we found that they are most likely aligned with serine, however there are minor variation. Serine peaks are observed at 969 cm$^{-1}$, 1010 cm$^{-1}$ and 1301 cm$^{-1}$, thus the appearing peaks at 938 cm$^{-1}$, 1164 cm$^{-1}$ are most probably originating from the neighbouring molecules (Asp-Ser-Ser-Pro-Asp-Ser-Ala-) or the citrate stabilizer. The appearance of the neighbouring molecules is either due to molecular diffusion or the hotspot partly covers more than one molecule of the peptide.

Oligonucleotides are also used in this method. The characteristics peaks of the four DNA nucleobases allow us to calculate the time of the molecular diffusion under the laser illumination. If the Raman signal is stable and only a certain number of molecules are visible, we conclude that the hotspot, the gold atomic structure and the nucleobase molecules are static. If additional nucleobases appear periodically, the measurement/technique can be considered dynamic and suitable for DNA or peptide sequencing.

**Nucleobase identification**

| References | [15] | [16] | [17] | [7] | [18] | Our assessment |
|---|---|---|---|---|---|---|
| **Adenine** | 735 | 732 | 735 | 735-737, 1467-1492 | 680, 720 | **735± 5** |
| **Cytosine** | 802 | 796 | 800 | 799-801, 1235-1270 | - | **802± 5** |
| **Guanine** | 651 | 661 | 660 | 954-958, 1545-1554 | - | **655± 5** |
| **Thymine** | 781 | 795 | 780 | 778-782, 1366-1373 | 740, 880 | **781± 5** |

Table 1: Characteristic Raman shifts of DNA nucleobases

The particle-in-well system was further used to detect the oligonucleotide (CCCATTTG) fragment. The collected Raman spectra was pre-processed before peak assignment. The collected Raman spectra was pre-processed before peak assignment. The four nucleobase peaks were assigned as follows based on the literature: adenine at 735 cm$^{-1}$, Cytosine at 802 cm$^{-1}$, Guanine at 651 cm$^{-1}$, and Thymine at 781 cm$^{-1}$ as shown in Table 1. Compared to multimolecular detection, single molecule peaks of nucleobases exhibit peak shifting, sometimes up to 15 cm$^{-1}$. We set up four Raman shift ranges corresponding four nucleobases. In our peak assignment, the maximum shift of 10 units of (± 5) from the most characteristic nucleobase peak was observed (Table 1). It is worth noting that certain characteristic in-plane ring breathing modes can be significantly suppressed due to the flat-lying molecule configuration under the hotspot, meaning that the single molecule configuration under the hotspot can enhance or even completely diminish the strongest peaks. [18] Apart from peak shifting, other common phenomenon is the peak drifting. Due to the molecule vibration, the single molecule peak doesn't provide a constant, sharp, and stable peak position, but exhibits dynamic movement. In results drifting can occur.

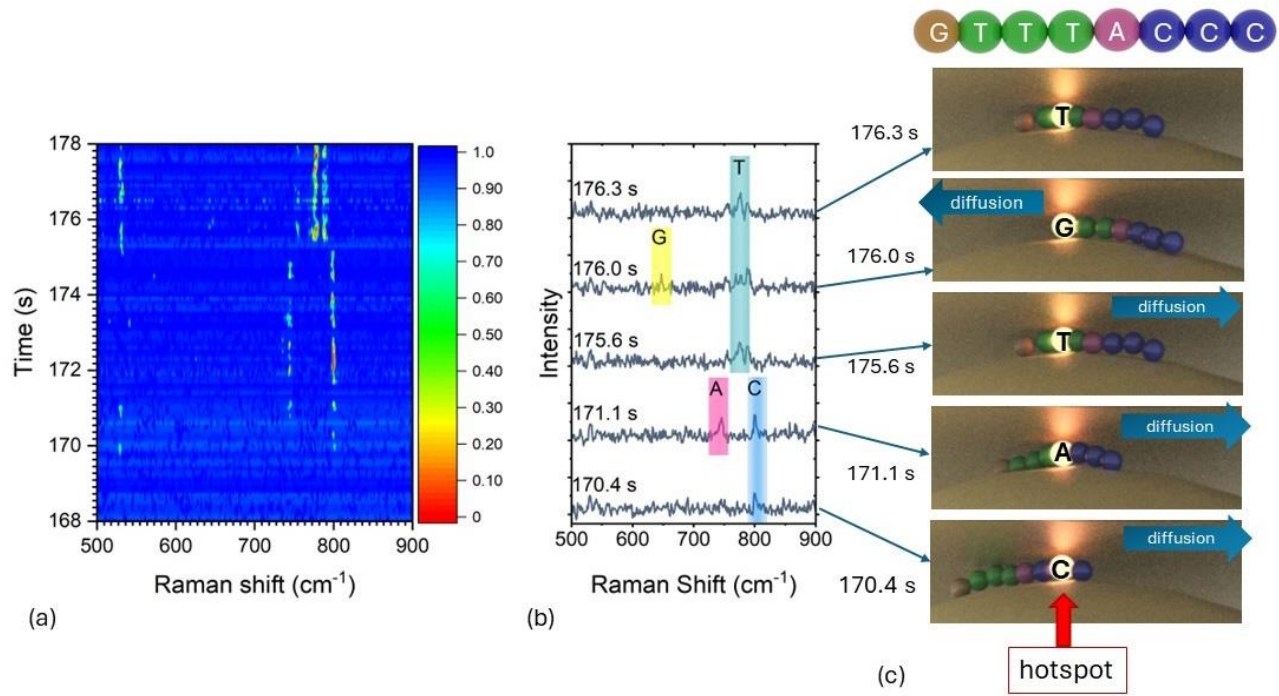

Figure 4: Discrimination of single nucleobases in oligonucleotides **a** Raman spectral map of CCCATTTG oligonucleotide **b** Peak assignment of nucleobases **c** Schematic of molecule diffusion under the plasmonic hotspot

The oligonucleotide sequence CCCATTTG was detected in the particle-in-well sensor. Based on 6 seconds of Raman spectra, the molecular diffusion rate is 1 second per nucleobase. Initially, the cytosine peak appears at 170.4 s, followed by adenine and cytosine together. In the next second, peaks for thymine, guanin and thymine appear sequentially again. This indicates that the molecular movement is neither linear nor restricted to a single direction.

**Conclusion**

To summarize, we implemented nanoparticle entrapment into the nanowell array (particle-in-well) system, that provides strong plasmonic enhancement for low concentration SERS detection. We demonstrated single molecule level detection by adsorbing a peptide chain on the gold nanoparticles. The observation of the building blocks of the peptide chain, valine and serine prove that the plasmonic hotspot is covering one molecule at the time, capable of single molecule detection. We further demonstrated nucleobase identification and monitoring of molecule diffusion of the oligonucleotide (CCCATTTG) on the nanoparticle surface. The SERS signals of the molecule diffusion could be used for single-molecule sequencing.

**Acknowledgements**

This research receives support from Academy Research Fellow project: TwoPoreProSeq (project number 347652), Biocenter Oulu emerging project (DigiRaman) and DigiHealth project (project number 326291), a strategic profiling project at the University of Oulu that is supported by the Academy of Finland and the University of Oulu.

**Experimental section**

**Materials**. Non-functionalized 50 nm gold nanoparticles (AuNPs) from Sigma (753645-25ML, concentration of $3.5 \times 10^{10}$ particles/mL) in 0.1 mM PBS citrate stbliized, reactant free. The synthetic peptides (Hypoxia-Inducible Factor-1α (HIF-1α) and fetuin-A (FETUA) were ordered from Biomatik. Dry DNA oligos (CCCATTTG, 4C4T4A4G, TAACTGGC, AAAAAAA, AAATAAA, AAACAAA, AAAGAAA) were purchased from Sigma. Silicon wafers with 100 nm $Si_3N_4$ membranes coated on the surface were purchased from MicroChemicals GmbH (WNA4 0525 155B 1314 S102, prime Si+ $Si_3N_4$ 4 inch).

**Fabrication of the nanowell arrays by FIB**

The nanowells that are fabricated by Focused Ion Beam (FIB). The substrate material is: 100 nm gold (sputtered) on a purchased wafer containing 100 nm $Si_3N_4$ on p-type Si 500μm. The nanowells were fabricated by using FIB and the estimated depth of the wells are ›100 nm in order to penetrate both the c (100 nm) and gold (100 nm) layer.

**Raman measurements**

SERS (ThermoFisher DXR2xi Raman Imaging Microscope) measurements on Particle-in-Well were carried out on selected nanoarrays with the following parameters: 50× objective, λ = 785 nm, laser power 18-22 mW, exposure time 0.1 s, full range resolution grid, 50 μm slit.